\documentclass[11pt]{article}

\usepackage{float}

\usepackage{tikz}
\usetikzlibrary{shapes,arrows} 

\usepackage[label font=it]{subfig}
\usepackage{caption}
\usepackage{multirow,tabularx}

\newsavebox{\twofigures}
\usepackage[export]{adjustbox}
\usepackage{calc}
\usepackage[version=4]{mhchem}
\usepackage{siunitx}
\usepackage{longtable,tabularx}
\usepackage{tabularray}
\usepackage{lmodern,pdftexcmds,amsmath}

\usepackage[title]{appendix}
\usepackage{titlesec}
\titleformat{\paragraph}
{\normalfont\normalsize}{\theparagraph}{1em}{}
\titlespacing*{\paragraph}
{0pt}{3.25ex plus 1ex minus .2ex}{1.5ex plus .2ex}
\counterwithin*{equation}{section}
\DeclareMathAlphabet{\mathsfbi}{OT1}{\sfdefault}{bx}{sl}
\DeclareMathVersion{sfletters}
\SetSymbolFont{letters}{sfletters}{OML}{ntxsfmi}{b}{it}

\makeatletter
\newcommand{\mathbfsbilow}[1]{%
  \text{\mathversion{sfletters}$\m@th#1$}%
}

\DeclareRobustCommand{\tensor}[1]{%
  \begingroup
  \ifcat\noexpand #1\relax
    \edef\greek@test{\detokenize{#1}}%
    \edef\greek@test{\expandafter\@cdr\greek@test\@nil}%
    \edef\greek@test{\expandafter\@car\greek@test\@nil}%
    \edef\x{\the\lccode\expandafter`\greek@test}%
    \edef\y{\number\expandafter`\greek@test}%
    \ifnum\x=\y\relax
      \mathbfsbilow{#1}%
    \else
      \mathsfbi{#1}%
    \fi
  \else
    \mathsfbi{#1}%
  \fi
  \endgroup
}
\makeatother

\usepackage{orcidlink}

\usepackage{tabularx, caption}
 \makeatletter
 \newcommand*{\compress}{\@minipagetrue}
 \makeatother

\newsavebox{\bigimage}

\usepackage{graphicx}
\usepackage{newtxtext}
\usepackage{newtxmath}
\usepackage{natbib}
\usepackage{hyperref}

\usepackage{algorithm}
\usepackage[noend]{algpseudocode}

\usepackage{etoolbox} 
\makeatletter
\patchcmd{\NAT@citex}
  {\@citea\NAT@hyper@{%
     \NAT@nmfmt{\NAT@nm}%
     \hyper@natlinkbreak{\NAT@aysep\NAT@spacechar}{\@citeb\@extra@b@citeb}%
     \NAT@date}}
  {\@citea\NAT@nmfmt{\NAT@nm}%
   \NAT@aysep\NAT@spacechar\NAT@hyper@{\NAT@date}}{}{}

\patchcmd{\NAT@citex}
  {\@citea\NAT@hyper@{%
     \NAT@nmfmt{\NAT@nm}%
     \hyper@natlinkbreak{\NAT@spacechar\NAT@@open\if*#1*\else#1\NAT@spacechar\fi}%
       {\@citeb\@extra@b@citeb}%
     \NAT@date}}
  {\@citea\NAT@nmfmt{\NAT@nm}%
   \NAT@spacechar\NAT@@open\if*#1*\else#1\NAT@spacechar\fi\NAT@hyper@{\NAT@date}}
  {}{}
\makeatother

\hypersetup{
    colorlinks = true,
     citecolor  = blue,
    allcolors=blue,
}

\newcommand{\RomanNumeralCaps}[1]

\usepackage{multirow,tabularx}

\usepackage[export]{adjustbox}
\usepackage{calc}
\usepackage[version=4]{mhchem}
\usepackage{siunitx}
\usepackage{longtable,tabularx}
\usepackage{tabularray}
 \UseTblrLibrary{amsmath}
 \usepackage{lmodern,pdftexcmds,amsmath}
\usepackage{booktabs}
\usepackage{tabularx}
\usepackage{array} 
\usepackage[noblocks]{authblk}
\usepackage[margin=1in]{geometry}
\usepackage[title]{appendix}

\newcommand\affiliation[1]{\gdef\@affiliation{\let\aff\aff@inst#1}}
\gdef\@affiliation{}
\def\email#1{Email address for correspondence: #1}
\def\aff#1{\ignorespaces\textsuperscript{#1}}
\def\corresp#1{\unskip\thanks{#1}}
\numberwithin{equation}{section}

\renewenvironment{abstract}
{\begin{quote}
\noindent \rule{\linewidth}{.5pt}\par{\bfseries \abstractname.}}
{\medskip\noindent \rule{\linewidth}{.5pt}
\end{quote}
}

\hypersetup{
   colorlinks=true,
   linkcolor={blue!100!black},
   citecolor={blue!100!black},
   plainpages=false,
}

\title{\bf Learning Differentiable Weak-Form Corrections to Accelerate Finite Element Simulations}

\author[1]{\bf Junoh Jung\corresp{\email{jjung@anl.gov}}}
\author[1]{\bf Emil Constantinescu}

\affil[1]{\normalsize Mathematics and Computer Science Division, Argonne National Laboratory, Lemont 60439 IL, USA \vspace{-1cm}}

\date{}
\begin{document}
\maketitle

\begin{abstract}
We present a differentiable weak-form learning approach for accelerating finite element simulations. Rather than introducing black-box source terms in the strong form of the governing equations, we augment the momentum equation directly in the variational (weak) form with parameterized bilinear operators. The coefficients of these operators are learned from high-resolution simulations so that unresolved small-scale dynamics can be represented on coarse grids. Applying the correction at the weak-form level aligns the learned model with the finite element discretization, preserving key numerical structure and better respecting the fundamental properties of incompressible flow. In the same setting, the approach yields solutions that are more accurate and more stable over long time horizons than comparable strong-form corrections. We implement the proposed method in the Firedrake finite element solver and evaluate it on benchmark problems, including the one-dimensional convection–diffusion equation and the two-dimensional incompressible Navier-Stokes equations. End-to-end differentiable training is enabled by coupling PyTorch with the Firedrake adjoint framework. Across these tests, the learned variational operators improve long-term accuracy while reducing computational cost. Overall, our results suggest that weak-form learning provides a principled, structure-preserving route to accurate and stable coarse-grid simulations of incompressible flows. \\  
\end{abstract}


\section{Introduction}
In the era of artificial intelligence, recent advances in machine learning (ML) have opened complementary avenues for fluid simulation. Purely data‑driven surrogates—ranging from physics‑informed neural networks (PINNs) to operator‑learning architectures and mesh‑graph networks—can emulate partial differential equation (PDE) solvers and deliver impressive speedups. However, ensuring long‑horizon stability, conservation, and generalization to out‑of‑distribution regimes remains challenging for end‑to‑end modeling. Representative contributions include the review by~\citet{Brunton2020} on machine learning for fluid mechanics, the physics-informed neural network (PINN) framework introduced by \citet{Raissi2019}, operator-learning approaches such as the Fourier neural operator (FNO) \citep{LiICLR2021FNO}, graph-based surrogate modeling exemplified by MeshGraphNets \citep{PfaffICLR2021}, and data-driven global weather forecasting with FourCastNet \citep{Pathak2022FourCastNet}.\\
\indent A promising alternative is hybrid physics–ML modeling, in which learned components are embedded within established PDE solvers to enhance accuracy and efficiency while preserving numerical structure. Prior efforts span learning discretizations \citep{BarSinai2019PNAS}, subgrid closures for large eddy simulation \citep{Beck2019JCP, Maulik2019JFM, Xie2020PRFluids}, Reynolds-averaged Navier--Stokes corrections with invariance \citep{Ling2016JFM}, and field‑inversion‑and‑ML approaches for model‑consistent augmentation \citep{ParishDuraisamy2016, Singh2017FIML}. Embedding ML inside the solver has been shown to improve accuracy at lower cost by leveraging the structure of the underlying numerical scheme. In compressible flows, recent work has accelerated discontinuous Galerkin (DG) solvers using learned correction terms with discrete updates \citep{de_lara_accelerating_2022, otmani_accelerating_2025}, and \citet{kang_differentiable_2025} further generalized the approach in DG and introduced continuous corrections via neural operators. Similar work was extended to a large-scale problem in a supercomputer setting~\citep{Jung2025HybridPhysicsML}. Despite this progress, few studies have examined learned corrections tailored to incompressible solvers on coarse grids, where maintaining stable long‑time behavior is critical.\\
\indent End-to-end learning in hybrid physics–ML computational fluid dynamics (CFD) refers to optimizing a learnable component (e.g., a closure or correction term) through the numerical solver against a global objective, such as mismatch in state variables, integral loads, or statistical quantities, so that training remains consistent with both the governing equations and their discrete time-marching implementation \citep{vinuesa_enhancing_2022,sanderse_scientific_2025}. When high-fidelity labels for the unclosed term are available (e.g., filtered subgrid stresses or eddy viscosity), a common alternative is a priori supervised training performed outside the CFD loop, which avoids the need to differentiate through the full solver trajectory \citep{sanderse_scientific_2025}. However, many practical settings  provide only indirect supervision (sparse measurements, pressure/velocity probes, forces, or long-time statistics). In this regime, end-to-end (embedded) training places the CFD solver in the loop and requires gradients of the objective with respect to ML parameters through the full discretized algorithm \citep{strofer_endtoend_2021,sirignano_dpm_2020,macart_embedded_2021,shankar_differentiable_2025}. These sensitivities can be obtained via discrete adjoint formulations and via automatic differentiation of a differentiable solver, both of which remain nontrivial for large-scale unsteady CFD \citep{strofer_endtoend_2021,bezgin_jaxfluids_2023,bezgin_jaxfluids2_2025}. Recent studies nevertheless demonstrate that solver-consistent training can materially improve stability and long-horizon accuracy of learned closures and hybrid architectures, including implementations on complex geometries and unstructured grids \citep{belbuteperes_combining_2020,list_learned_2022,kim_generalizable_2024}.\\
\indent Firedrake~\citep{FiredrakeUserManual} is an automated finite element framework for the solution of PDEs using solvers based on the Portable, Extensible Toolkit for Scientific Computation (PETSc)~\citep{PETSC2019}, which can provide a natural entry point for differentiable PDE programs at the level of weak forms \citep{rathgeber_firedrake_2016}. Its adjoint capability (firedrake.adjoint) builds on the pyadjoint approach~\citep{mitusch_dolfin_adjoint_2019}. Specifically, the forward model is annotated (“taped”), and then discrete tangent-linear/adjoint models are derived and solved automatically, enabling scalable gradient evaluation for PDE-constrained optimization and inverse problems \citep{farrell_automated_2013,mitusch_dolfin_adjoint_2019,funke_framework_2013}. This differentiable foundation has been exercised extensively in gradient-based inference workflows built on Firedrake, including recent work that introduces a composable and differentiable point-evaluation abstraction for consistent data assimilation in Firedrake/Icepack \citep{nixonhill_point_assimilation_2024}. 
Firedrake’s differentiable programming story also has been extended explicitly toward hybrid physics–ML training. A foreign-function/external-operator interface was introduced so that operators not representable in pure United Form Language vector calculus, such as neural network closures implemented in PyTorch, can appear directly inside variational forms while remaining compatible with assembly and adjoint differentiation \citep{bouziani_escaping_ufl_2021}. Building on these ideas, subsequent interfaces couple Firedrake with mainstream ML components by wrapping Firedrake computations as custom operators for PyTorch, thereby enabling end-to-end training of coupled PDE-neural models using standard deep-learning optimizers while still leveraging Firedrake’s code-generated performance and adjoint machinery \citep{bouziani_pytorch_firedrake_2023,bouziani_pde_ml_barrier_2024}.\\
\indent Beyond strong‑form correction, a growing body of work in scientific ML emphasizes variational/weak‑form learning, coupling data‑driven components to the same integrated, test‑function–weighted equations that underlie Galerkin finite elements and many projection‑based reduced‑order models. Recent efforts have used weak forms as the interface for learned correction/closure operators: in variational multiscale  reduced-order models, neural networks learn the influence of unresolved scales while preserving Galerkin structure and energy/consistency properties \citep{mou_data_driven_2021,ahmed_physics_guided_2023,dar_artificial_2023,ivagnes_pressure_2023,koc_residual_based_2025}. Related work embeds learning into stabilized Petrov--Galerkin formulations by learning stabilization parameters or residual-based terms that are naturally expressed at the element/weak level \citep{tassi_supg_2023}. Parallel threads train models by minimizing weak residuals to target weak solutions more directly \citep{zang_weak_2020,chen_friedrichs_2023,de_ryck_wpinns_2024}.\\
\indent We present a weak-form correction approach for hybrid physics–ML modeling on coarse grids for an incompressible finite element solver. We augment the momentum equation with parameterized bilinear contributions whose coefficients are trained from high-resolution solutions projected onto coarse grids. Learning in the weak form couples the correction to the PDE operator, enabling structure-preserving designs that respect numerical stability in a finite element solver. We demonstrate the approach on two test cases: the 1D convection–diffusion equation and the 2D Navier–Stokes equations.\\
\indent The remainder of this paper is organized as follows. Section 2 describes the simulation setup in Firedrake and the datasets used. Section 3 details the methodology and the end-to-end training implementation in Firedrake. Section 4 presents results for the 1D convection–diffusion equation and 2D flow past a cylinder. Section 5 concludes with a summary and directions for future work.

\section{Simulation and datasets}
This section outlines the simulation settings and datasets used in our study. We consider two examples: the one-dimensional convection–diffusion equation and the two-dimensional incompressible Navier–Stokes equations for flow past a cylinder.
\subsection{Simulation}
We use the finite-element-based Firedrake~\citep{FiredrakeUserManual} framework built on PETSc~\citep{PETSC2019} to formulate both problems in weak form and to solve the linear and nonlinear systems using PETSc’s SNES/KSP solvers with appropriate preconditioning for the linearized subproblems. We couple Firedrake with PyTorch and employ differentiable programming tools that provide end-to-end derivatives of simulation-based objectives with respect to physical parameters, initial/boundary conditions, and learnable closures. Time-dependent cases are advanced with a Crank--Nicolson (CN) scheme. At each step, we solve the discrete residual to a fixed tolerance and record the fields required to construct the training/test datasets described below.
\subsection{One-dimensional convection--diffusion equation}\label{subsec:1d-convdiff}
\begin{figure}[t]
  \centering
   \includegraphics[scale=0.6,width=0.6\textwidth]{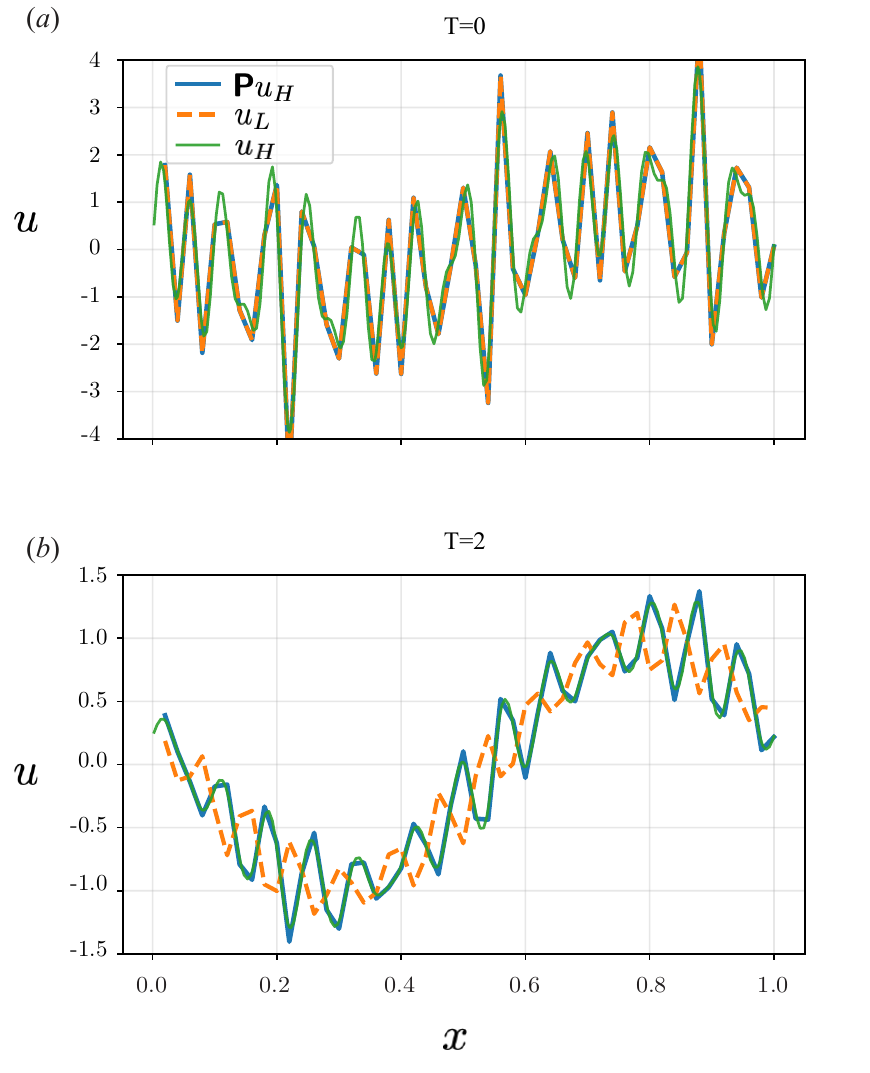}
  \caption{ 1D CONVECTION-DIFFUSION EQUATION: (\textit{a}) INITIAL CONDITION AND (\textit{b}) SOLUTION AT $T=2$.}\label{fig_1d_initial}
\end{figure}

\begin{figure}[!htp]
  \centering
   \includegraphics[scale=0.6,width=0.6\textwidth]{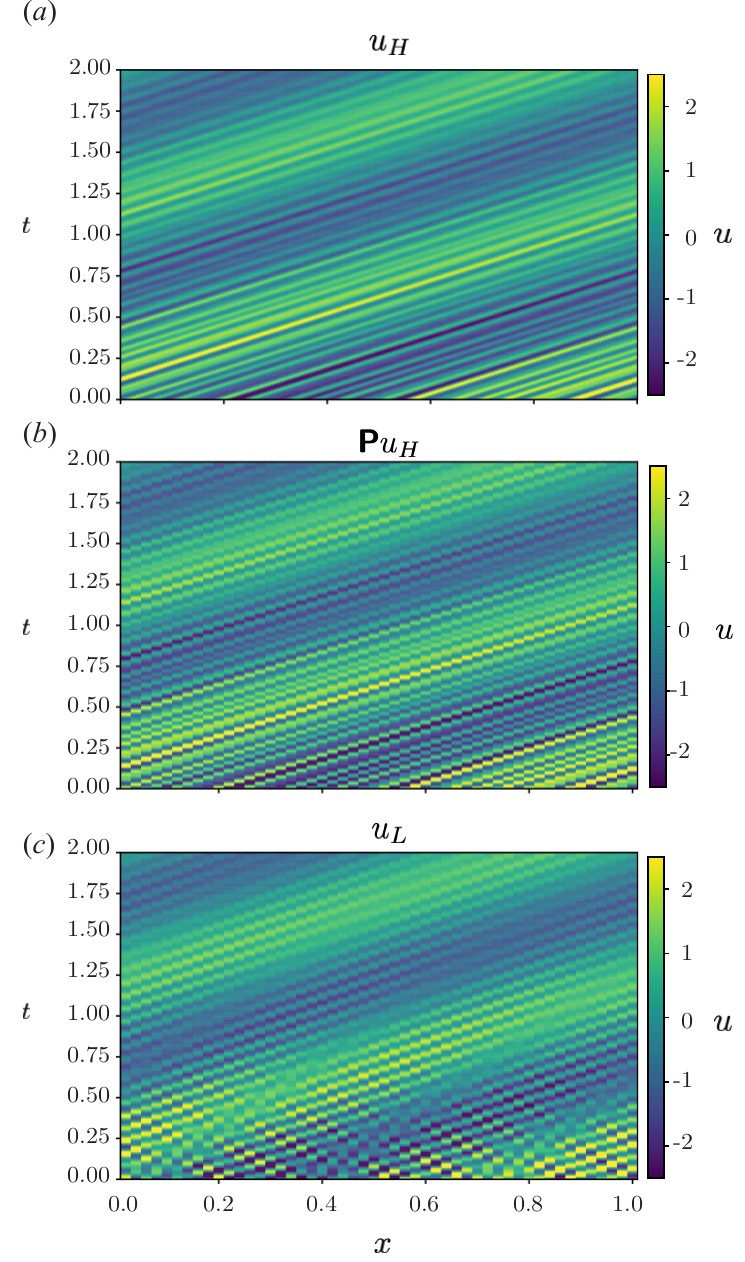}
  \caption{ SPACE-TIME FIELDS FOR THE 1D CONVECTION-DIFFUSION EQUATION. (\textit{a}) HIGH-RESOLUTION SOLUTION $u_{H}$. (\textit{b}) PROJECTED FIELD $\mathsfbi{P}u_{H}$. (\textit{c}) LOW-RESOLUTION SOLUTION $u_{L}$.}\label{Fig_1d_spaceandtimefield}
\end{figure}

We first study the propagation of a multi-frequency initial signal governed by the 1D convection--diffusion equation: 
\begin{equation}  \frac{\partial u}{\partial t} + a \frac{\partial u}{\partial x}  = \nu \frac{\partial^2 u}{\partial x^2},  \qquad x \in [0,1], \quad t \in [0,T],  \label{eq:1Dconvdiff}
\end{equation}
where $u(x,t)$ denotes the scalar field, $a$ is the convection coefficient, and $\nu$ is the diffusion coefficient. In all tests we fix $a = 1$ and $\nu = 0.0001$ in Eq.~\eqref{eq:1Dconvdiff}, and we impose periodic boundary conditions at $x=0$ and $x=1$. The initial condition is prescribed as
\begin{equation} u(t=0,x)=4\sum_{i=1}^{4}\sin\!\big(2\pi \alpha_i(x-\phi)\big), \label{eq:init_cond} \end{equation} 
with $\alpha_i=\{4,6,7,20\}$ and a phase shift $\phi\sim\mathcal{U}(0,1)$, as shown in Figure~\ref{fig_1d_initial}(\textit{a}). For each sampled $\phi$, Eq.~\eqref{eq:1Dconvdiff} is advanced in time by using a CN scheme. To generate reference data, we run simulations for $100$ randomly sampled phases $\phi$ over $t\in[0,2]$ on a mesh of $50$ elements using a high-order continuous Galerkin  discretization in Firedrake. The last step solution at $T=2$ is shown in Figure~\ref{fig_1d_initial}(\textit{b}). The resulting spatiotemporal high-fidelity solutions are denoted by $u_H$ ($P=5$), yielding 250 degrees of freedom (DOFs), and are shown in Figure~\ref{Fig_1d_spaceandtimefield}(\textit{a}). The coarse field is computed via the $L_2(\Omega)$ (Galerkin) projection of the high-resolution solution $u_H \in V_H$ onto the low-resolution space $V_L$. Specifically, $P u_H \in V_L$ is defined by \begin{equation} \bigl(Pu_H, v\bigr)_{L_2(\Omega)} = \bigl(u_H, v\bigr)_{L_2(\Omega)} \qquad \forall v \in V_L, \end{equation} where the $L_2(\Omega)$ inner product is $ (a,b)_{L_2(\Omega)} := \int_{\Omega} a(x)\, b(x)\, dx$. The coarse field $\mathsfbi{P}u_H$ is shown in Figure~\ref{Fig_1d_spaceandtimefield}(\textit{b}). These projected fields are used to construct the training and test datasets. We use the projected trajectory $\mathsfbi{P}u_H$ as the learning target. Training data span $t\in[0,2]$, and testing uses a separate $t\in[0,5]$ segment with a different phase. This phase-shifted test set is then used to assess longer-horizon predictive performance. For comparison, the corresponding low-resolution simulation ($P=1$) is shown in Figure~\ref{Fig_1d_spaceandtimefield}(\textit{c}). \\ 
\subsection{Two-dimensional flow past a cylinder}
\begin{figure}[t]
  \centering
   \includegraphics[scale=0.9,width=0.9\textwidth]{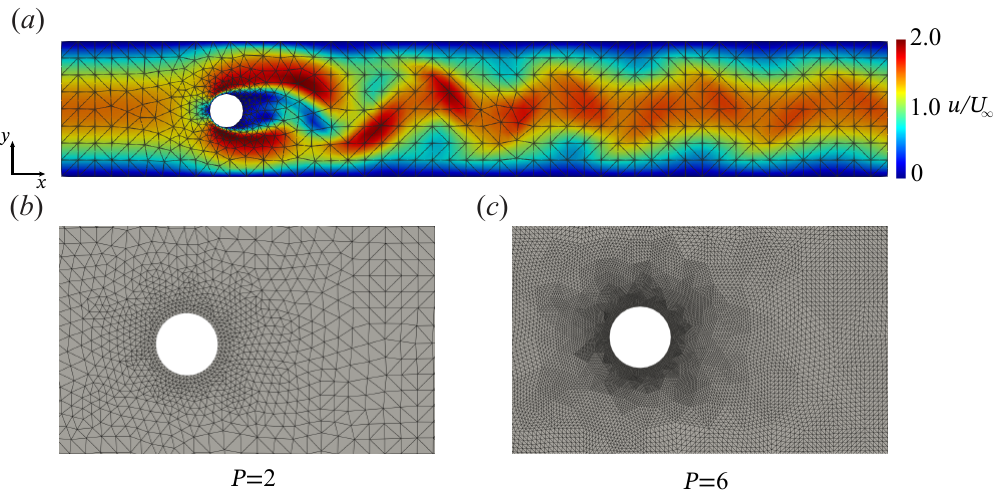}
  \caption{ 2D FLOW PAST A CYLINDER: (\textit{a}) COMPUTATIONAL MESH COLORED BY THE INSTANTANEOUS VELOCITY MAGNITUDE (ONLY ELEMENT EDGES ARE SHOWN). (\textit{b}) HIGH-RESOLUTION GRID ($p_v,p_p = 6,5$) WITH ZOOMED-IN MESH INSETS NEAR THE CYLINDER. (\textit{c}) LOW-RESOLUTION GRID ($p_v,p_p = 2,1$).}\label{Fig_cylin_grid}
\end{figure}
We simulate two-dimensional flow past a circular cylinder in a channel at Reynolds number 100 using Firedrake \citep{FiredrakeUserManual} on the standard benchmark geometry \citep{hirschel_benchmark_1996}. 
The incompressible Navier--Stokes equations on a bounded domain
$\Omega\subset\mathbb{R}^2$ with boundary $\partial\Omega$ over $t\in(0,T]$ are
\begin{equation}
\begin{aligned}
\partial_t \mathbf{u}
+ (\mathbf{u}\cdot\nabla)\mathbf{u}
+ \nabla p
- \nu \Delta \mathbf{u}
&= 0
\quad \text{in } \Omega,
\\
\nabla\cdot \mathbf{u}
&= 0
\quad \text{in } \Omega .
\end{aligned}
\label{eq_NSeqn}
\end{equation}
We adopt the standard spaces $V=[H^1(\Omega)]^2$ and $Q=L^2(\Omega)$, with finite-dimensional subspaces $V_h\subset V$ and $Q_h\subset Q$. Eq. ~\eqref{eq_NSeqn} is solved with no-slip conditions on the cylinder and channel walls, a laminar parabolic inflow at \(\Gamma_I\),\[u_{\text{in}}(y) = \frac{4V_{\max}\,y(0.41-y)}{0.41^2}, \quad v_{\text{in}}(y)=0 \quad (V_{\max}=1.5),\]and a traction-free outflow at \(\Gamma_O\). Baseline parameters are $\nu=0.001$, $\Delta t=0.01$, $T_0=0$, and $T_{\mathrm{end}}=53$. The training set spans $300\,\Delta t$, and the test set spans $5000\,\Delta t$. The computational domain is \(\Omega = [0,\,2.5] \times [0,\,0.41]\) with a circular cylinder of radius \(0.05\) embedded in the channel, shown in Figure~\ref{Fig_cylin_grid}. We discretize in space using inf–sup-stable Taylor--Hood elements. For the high-fidelity reference solution, we employ polynomial degrees \((p_v, p_p)=(6,5)\), resulting in 40,884 velocity DOFs and 14,255 pressure DOFs. For the coarse-resolution runs we use \((2,1)\), which yields 4,732 velocity DOFs and 627 pressure DOFs. Time integration is performed with a CN scheme at each time step.

\section{Methodology}
We begin by briefly reviewing strong-form correction strategies used in prior work \citep{de_lara_accelerating_2022,Kang2023Learning,otmani_accelerating_2025,kang_differentiable_2025,Jung2025HybridPhysicsML}. Next, we motivate a weak-form perspective using the two-dimensional incompressible Navier--Stokes equations as a representative example. Then, we present the proposed weak-form correction approach. Finally, we describe the training objective and the differentiable learning pipeline, which combines discrete adjoints in the Firedrake solver with backpropagation implemented in PyTorch.

\subsection{Strong-form correction}\label{subsec:methodology_st-form_correction}
In general, a strong-form correction augments the governing equation as \begin{equation} 
\frac{\partial \boldsymbol{u}}{\partial t} = \mathcal{F}\big(\boldsymbol{u}(t)\big) + \mathcal{N}_{\theta}(\boldsymbol{u}), \label{eq:strong_correction} 
\end{equation}
where $\mathcal{F}$ denotes the nonlinear operator defining the resolved dynamics. The term $\mathcal{N}_{\theta}(u)$ is a learned neural operator that acts as an additional source to account for unresolved (subgrid) effects on a coarse grid. In this setting, training is performed through the time-integration procedure, which requires a differentiable solver to enable multistep, unsupervised learning.\\ 
\indent In a finite element formulation, this correction appears in the weak form as the pointwise forcing term $(\mathcal{N}_{\theta}(\boldsymbol{u}), \boldsymbol{v})$. This contrasts with approaches based on bilinear-form (operator) corrections, which modify the discrete operator itself rather than adding an explicit source.
\subsection{Weak-form correction}\label{subsec:methodology_weak-form_correction}
The semi-discrete Galerkin formulation of the incompressible Navier--Stokes equations seeks $(\mathbf{u}_h(t),p_h(t))\in V_h\times Q_h$ such that for all $(\mathbf{v}_h,q_h)\in V_h\times Q_h$ and $t\in(0,T]$, 
\begin{equation}\label{eq:NS-weak} 
\begin{aligned} 
(\partial_t \mathbf{u}_h,\mathbf{v}_h)_\Omega &+ \nu\,(\nabla\mathbf{u}_h,\nabla\mathbf{v}_h)_\Omega + c_{\mathrm{skew}}(\mathbf{u}_{\mathrm{adv}};\mathbf{u}_h,\mathbf{v}_h) \\ &- (p_h,\nabla\!\cdot\!\mathbf{v}_h)_\Omega + (q_h,\nabla\!\cdot\!\mathbf{u}_h)_\Omega = 0. 
\end{aligned} 
\end{equation} 
Here, $\mathbf{u}_{\mathrm{adv}}$ denotes the advecting velocity (e.g., $\mathbf{u}_h$ for the fully nonlinear form or a lagged/extrapolated field in linearized Oseen-type updates). We adopt the skew-symmetric convection operator 
\begin{equation}\label{eq:cskew}
c_{\mathrm{skew}}(\mathbf{a};\mathbf{w},\mathbf{v})
=\tfrac12\Big((\mathbf{a}\!\cdot\!\nabla \mathbf{w},\mathbf{v})_\Omega
-(\mathbf{a}\!\cdot\!\nabla \mathbf{v},\mathbf{w})_\Omega\Big).
\end{equation} 
This form is skew-adjoint and therefore satisfies $c_{\mathrm{skew}}(\mathbf{a};\mathbf{w},\mathbf{w})=0$ for any $\mathbf{a}$, which prevents artificial kinetic energy production from the discrete convection term.\\
\indent To obtain a stable and physically interpretable correction mechanism within the finite element weak setting, we introduce three scalar coefficient fields $c_{\mathrm{adv}}(\mathbf{u}_{h})$, $\nu_t(\mathbf{u}_{h})$, and $\gamma(\mathbf{u}_{h})$. Each coefficient is represented in a low-order scalar finite element space $S_h$, so that $c_{\mathrm{adv}}(\mathbf{u}_{h}), \nu_t(\mathbf{u}_{h}), \gamma(\mathbf{u}_{h}) \in S_h$ and, in the discrete implementation, each field is encoded by a vector of $N_{\mathrm{coef}} = \text{dim} (S_h)$ DOFs. We constrain the coefficients to lie in the bounds $c_{\mathrm{adv}}(\mathbf{u}_{h}) \in [-c_{\max},c_{\max}]$, $\nu_t(\mathbf{u}_{h}) \in [0,\nu_{t,\max}]$, and $\gamma(\mathbf{u}_{h}) \in [0,\gamma_{\max}]$, which modulate the strength of advection ($c_{\mathrm{adv}}$), augment diffusion through an eddy viscosity ($\nu_t$), and provide grad-div stabilization ($\gamma$), respectively. The corrected semi-discrete problem consists in determining $(\mathbf{u}_h(t),p_h(t))\in V_h\times Q_h$ such that for all $(\mathbf{v}_h,q_h)\in V_h\times Q_h$, 
\begin{equation}\label{eq:NS-weakcorr} 
\begin{aligned} (\partial_t \mathbf{u}_h,\mathbf{v}_h)_\Omega &+ \big((\nu+\nu_t)\nabla\mathbf{u}_h,\nabla\mathbf{v}_h\big)_\Omega \\ &+ \big(1+c_{\mathrm{adv}}\big)\, c_{\mathrm{skew}}(\mathbf{u}_{\mathrm{adv}};\mathbf{u}_h,\mathbf{v}_h) \\ &- (p_h,\nabla\!\cdot\!\mathbf{v}_h)_\Omega + (q_h,\nabla\!\cdot\!\mathbf{u}_h)_\Omega \\ &+ \big(\gamma\,\nabla\!\cdot\!\mathbf{u}_h,\nabla\!\cdot\!\mathbf{v}_h\big)_\Omega + \varepsilon_p (p_h,q_h)_\Omega = 0 . 
\end{aligned} 
\end{equation} 
The small parameter $\varepsilon_p\ge 0$ is an optional pressure penalty used solely to regularize the mixed linear algebra for certain solver configurations. The corrections enter through bilinear forms assembled in the same finite element framework, which preserves sparsity and locality, maintains compatibility with existing solvers and preconditioners, and supports discrete-adjoint consistency.
\subsection{Training objective and end-to-end training via discrete adjoints and backpropagation}
We define the training objective and summarize the learning algorithm. Given snapshot data
$\{\mathbf{u}^n\}$, we define the per-step mismatch functional
\begin{align}
J^{n+1}(\theta)
= \frac{1}{2}\int_\Omega \left\|\tilde{\mathbf{u}}^{n+1}_{h,\theta} - \mathbf{u}^{n+1}\right\|^2\,d\Omega,
\label{eq:loss-onestep}
\end{align}
where $\tilde{\mathbf{u}}^{n+1}_{h,\theta}$ is the one-step prediction produced by the corrected solver.
Training uses random restart indices $s_0$ and rollouts of length $m$ time steps. For each rollout and
each substep $j=0,\dots,m-1$, we construct correction fields from the current state
$\mathbf{u}_h^{s_0+j}$, advance the corrected weak form~\eqref{eq:NS-weakcorr} to obtain
$\tilde{\mathbf{u}}^{s_0+j+1}_{h,\theta}$, and evaluate \eqref{eq:loss-onestep} against
$\mathbf{u}^{s_0+j+1}$. The total rollout objective is 
\begin{align}
J(\theta) = \sum_{j=0}^{m-1} J^{s_0+j+1}(\theta),
\label{eq:sum-loss}
\end{align}
and we differentiate $J$ through the entire $m$-step time-marching map. Algorithm~\ref{algorithm_training}
summarizes the procedure.

We compute discrete sensitivities of the rollout objective with respect to the correction fields using
algorithmic differentiation of the finite element solver (Firedrake adjoint). In particular, for each
substep we obtain $\frac{dJ}{dc_{\mathrm{adv}}}$, $\frac{dJ}{d\nu_{\mathrm{t}}}$, and $\frac{dJ}{d\gamma}$
in the coefficient dual space $S_h^\ast$. These adjoint sensitivities provide upstream gradients for the
neural parameterization, and the chain rule gives
\begin{equation}
\frac{dJ}{d\theta}
=
\sum_{j=0}^{m-1}
\underbrace{\left(\frac{\partial \boldsymbol{\eta}_{s_0+j}}{\partial \theta}\right)^\top}_{\substack{\text{PyTorch}}}
\underbrace{\frac{dJ}{d\boldsymbol{\eta}_{s_0+j}}}_{\substack{\text{Firedrake adjoint}}},
\label{eq:chainrule_rollout}
\end{equation}
where $\boldsymbol{\eta}$ denotes the learned coefficient fields for
$(c_{\mathrm{adv}},\nu_{\mathrm{t}},\gamma)$.

In implementation, we couple solver-adjoint gradients to PyTorch by seeding backpropagation on the
coefficient tensors with the adjoint-derived sensitivities. Parameters are updated with Adam. Gradient
clipping and learning-rate scheduling are applied to improve robustness.

\begin{algorithm}[t]
\caption{End-to-end training of weak-form correction (rollout length $m$)}
\label{algorithm_training}
\footnotesize
\begin{algorithmic} 
\State \textbf{Input:} snapshots $\{\mathbf{u}^n\}_{n=0}^{T}$, timestep $\Delta t$, rollout $m$, model $\mathcal{N}_\theta$
\Repeat
  \State Sample mini-batch $B$ of restart indices $s_0$; \textbf{zero grads}
  \For{$s_0 \in B$}
    \State $\mathbf{u}\gets \mathbf{u}^{s_0}$; $J\gets 0$
    \For{$j=0$ \textbf{to} $m-1$} \Comment{\scriptsize forward rollout (taped)}
      \State $\boldsymbol{\eta}_{s_0+j}\gets \mathcal{N}_\theta(\mathrm{vec}(\mathbf{u}))$
      \State $\tilde{\mathbf{u}}\gets$ solve \eqref{eq:NS-weakcorr} with $\boldsymbol{\eta}_{s_0+j}$
      \State $J \gets J + \tfrac12 \|\tilde{\mathbf{u}}-\mathbf{u}^{s_0+j+1}\|_{L^2}^2$
      \State $\mathbf{u}\gets \tilde{\mathbf{u}}$
    \EndFor
    \State $\left\{\dfrac{dJ}{d\boldsymbol{\eta}_{s_0+j}}\right\}_{j=0}^{m-1}\gets$
           discrete adjoint of taped rollout (Firedrake adjoint)
    \State Backpropagate on each $\boldsymbol{\eta}_{s_0+j}$ with $\dfrac{d\eta_{s_0+j}}{d\boldsymbol{\theta}}$ (PyTorch)
  \EndFor
\Until{converged}
\State \textbf{Return:} trained parameters $\theta^\ast$
\end{algorithmic}
\end{algorithm}

\section{Results}
This section reports results for two test cases that demonstrate the proposed weak-form correction approach. For each example, we compare weak-form and strong-form corrections in terms of (i) training performance (loss history), (ii) predictive accuracy relative to the projected high-resolution reference solution, and (iii) computational efficiency, quantified by speedup relative to the high-resolution solver and by time-averaged error over long rollouts, in other words, the effective prediction horizon. All experiments were run on an Apple Silicon Mac (M4) using PyTorch’s MPS GPU backend.
\subsection{One-dimensional convection--diffusion equation}
\begin{figure}[t]
  \centering
   \includegraphics[scale=0.5,width=0.5\textwidth]{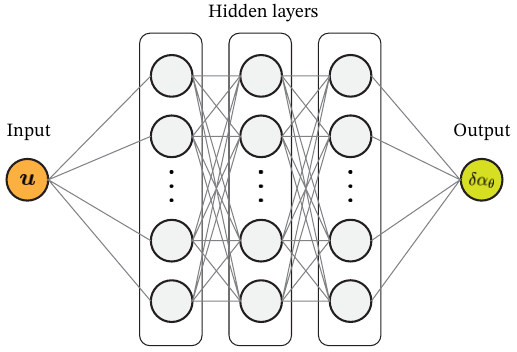}
  \caption{ARCHITECTURE OF CORRECTIVE OPERATOR FOR 1D CONVECTION-DIFFUSION EQUATION. }\label{Fig_architecture_1DConvDiff}
\end{figure}
\begin{figure}[t]
  \centering
   \includegraphics[scale=0.9,width=0.9\textwidth]{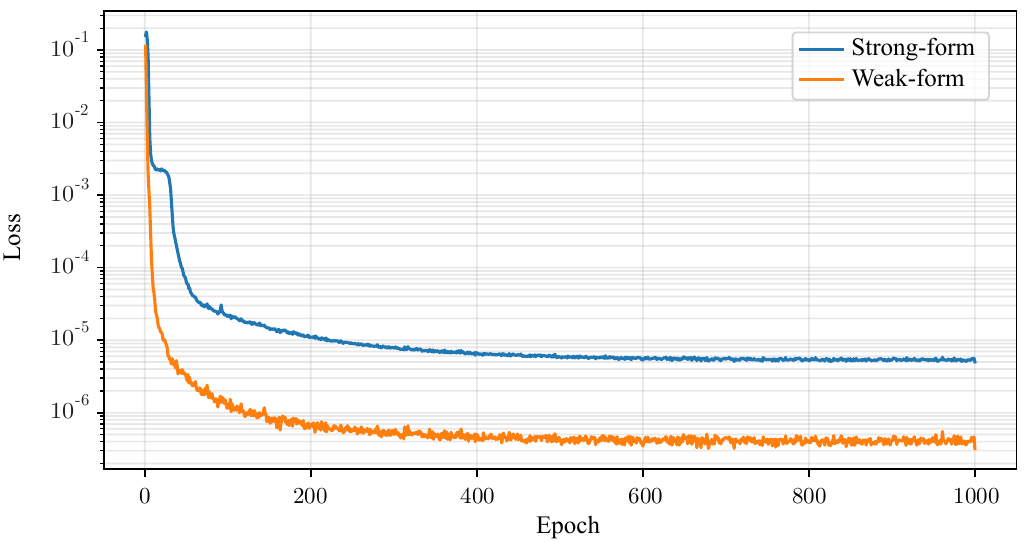}
  \caption{TRAINING LOSS VERSUS EPOCH FOR STRONG-FORM AND WEAK-FORM CORRECTIONS IN 1D CONVECTION-DIFFUSION EQUATION. }\label{Fig_Loss_1DConvDiff}
\end{figure}
We start from a semi-discrete weak formulation on the coarse finite element space and introduce neural corrections directly at the variational level. In the experiments reported here, we employ a bilinear weak-form correction. That is, we learn a correction field that enters through the term paired with the test-function derivative, \begin{equation} 
\delta \boldsymbol{\alpha}_\theta = \mathcal{N}_\theta\!\left(\boldsymbol{u}_h\right), 
\end{equation} 
where $\mathcal{N}_\theta$ is a neural network evaluated on the nodal degree of freedom vector of the coarse solution $\boldsymbol{u}_h$. The corrected weak formulation seeks $\boldsymbol{u}_h(t)\in V_h$ such that, for all $\boldsymbol{v}_h\in V_h$ and $t\in(0,T]$, 
\begin{equation} 
\begin{aligned} 
(\partial_t \boldsymbol{u}_h, \boldsymbol{v}_h)_\Omega &+ (a\,\boldsymbol{u}_{h,x}, \boldsymbol{v}_h)_\Omega + (\nu\,\boldsymbol{u}_{h,x}, \boldsymbol{v}_{h,x})_\Omega + (\delta \boldsymbol{\alpha}_\theta, \boldsymbol{v}_{h,x})_\Omega = 0 . 
\end{aligned} \label{eq:corrected_weakform_convdiff_fluxonly} \end{equation} Here $(\delta \boldsymbol{\alpha}_\theta, \boldsymbol{v}_{h,x})_\Omega$ acts as a learned correction to the resolved transport-diffusion balance through a weak flux-like contribution. Since periodic boundary conditions are imposed at $x=0$ and $x=1$, no boundary integrals arise in Eq.~\eqref{eq:corrected_weakform_convdiff_fluxonly}. When the baseline coarse model is sufficient, training drives $\delta \alpha_\theta\approx 0$, recovering the standard Galerkin method.\\ 
\indent For training, we minimize a multistep rollout loss over a horizon of length $m=20$. Mini-batches are constructed by randomly sampling short subsequences from the stored projected reference trajectories $\mathsfbi{P}u_H$. For each sampled subsequence, the model is rolled out for $m$ steps, and the parameters are updated by minimizing the mean-squared error between the predicted rollout and the reference subsequence over the full horizon. Time integration during both training and evaluation is performed using a CN scheme. The correction field $\delta\boldsymbol{\alpha}_\theta$ is parameterized by a multilayer perceptron  with three ReLU-activated hidden layers of widths $(128,128,128)$, as shown in Figure~\ref{Fig_architecture_1DConvDiff}. We train for 1,000 epochs with batch size $B=10$ and 10 batches per epoch. The learning rate is initialized at $10^{-3}$ and decayed multiplicatively by a factor of $0.99$ per epoch. Both the strong-form and weak-form models are trained on the same dataset using the same rollout loss. They use identical network architectures with the same number of parameters. The only difference is where the learned correction is applied, either as a strong-form term or as a weak-form variational contribution.\\ 
\indent As shown in Figure~\ref{Fig_Loss_1DConvDiff}, the training loss decreases rapidly in the initial epochs and then gradually levels off. Small fluctuations in the loss are expected due to random mini-batch sampling. Over the full training window, the weak-form model consistently achieves a lower loss than the strong-form model, consistent with weak-form correction providing a numerically structure-preserving formulation that is less sensitive to pointwise high-frequency residuals.\\
\begin{figure}[t]
  \centering
   \includegraphics[scale=0.8,width=0.8\textwidth]{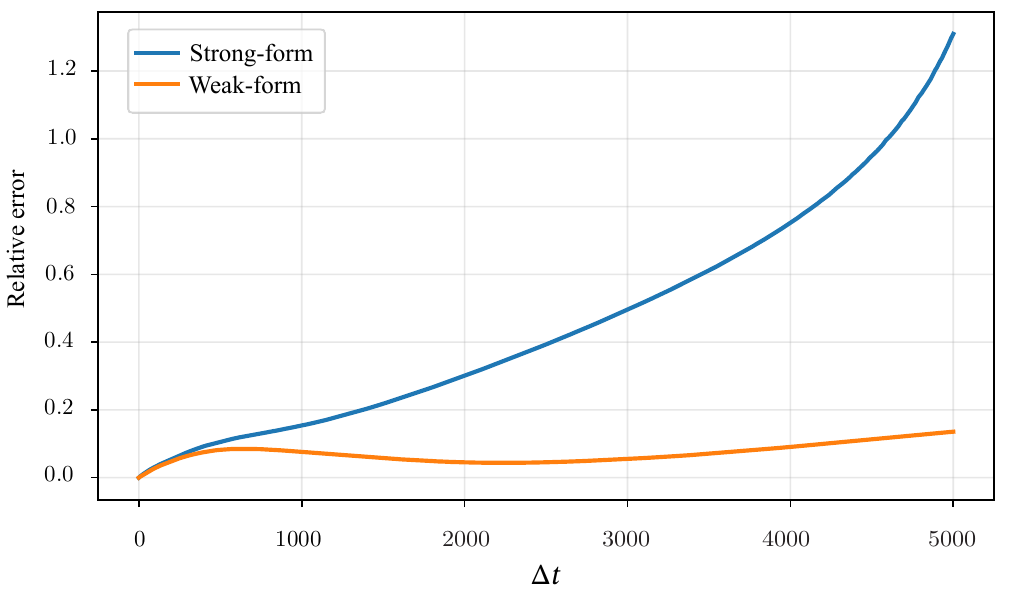}
  \caption{TIME EVOLUTION OF THE RELATIVE ROLLOUT ERROR FOR THE 1D CONVECTION-DIFFUSION EQUATION USING STRONG-FORM AND WEAK-FORM NEURAL CORRECTIVE OPERATORS.}\label{Fig_relativeError_1Dconv}
\end{figure}
\begin{figure}[t]
  \centering
   \includegraphics[scale=0.8,width=0.8\textwidth]{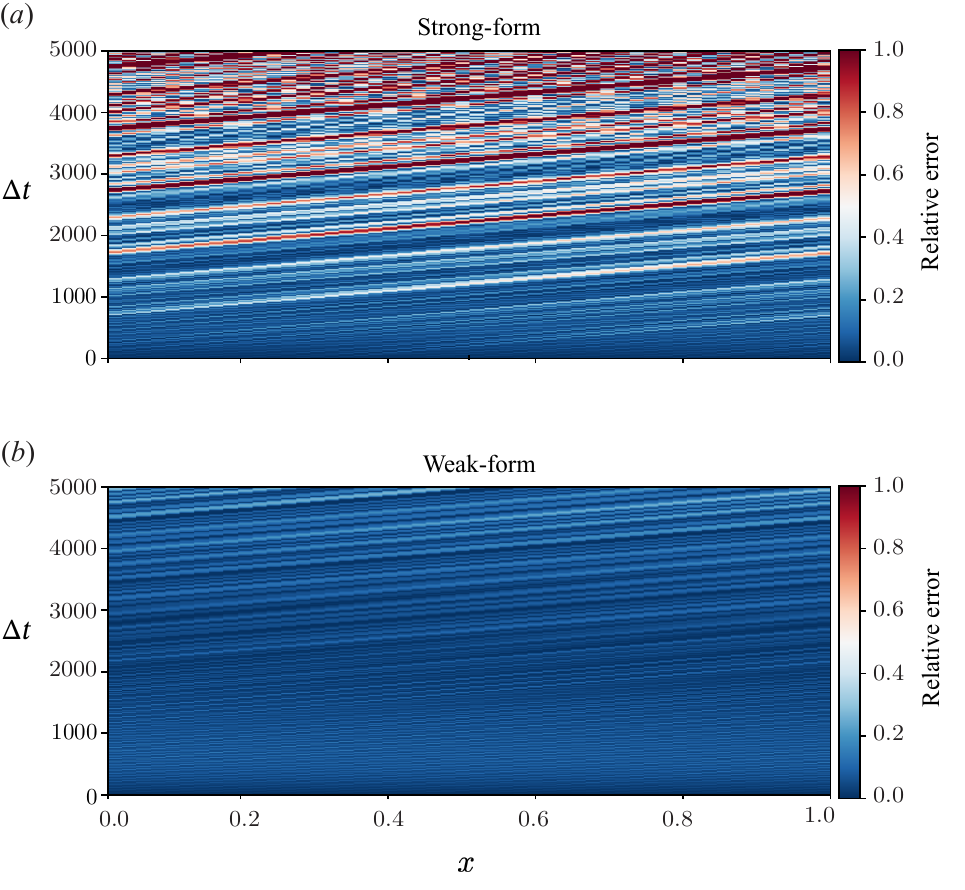}
  \caption{RELATIVE ERROR FIELDS IN SPACE-TIME FOR THE 1D CONVECTION-DIFFUSION EQUATION, COMPARING (\textit{a}) STRONG-FORM AND (\textit{b}) WEAK-FORM CORRECTIVE OPERATORS.}\label{Fig_Error_field_1Dconv}
\end{figure}
\indent Figure~\ref{Fig_relativeError_1Dconv} evaluates long-horizon rollout accuracy using the normalized relative error \begin{equation}\label{eq:relativeerror} e(t)=\frac{\|\tilde{u}(t)-\mathsfbi{P}u_{H}(t)\|_{2}}{\|\mathsfbi{P}u_{H}(t)\|_{2}}, \end{equation} where $\tilde{u}$ denotes the model prediction and $\mathsfbi{P}u_H$ is the projected reference trajectory on the coarse space. Across the full rollout interval ($t\in[0,5000\Delta t]$), the weak-form correction achieves a lower relative error than does the strong-form correction. Moreover, the weak-form model exhibits a slower growth rate of $e(t)$, particularly at later times, indicating improved robustness under iterative time stepping when error accumulation typically becomes dominant. In contrast, the strong-form correction shows a larger drift and faster error growth, reflecting reduced stability in long-horizon prediction. The space-time relative-error maps in Figure~\ref{Fig_Error_field_1Dconv} further clarify these trends. The strong-form correction produces higher error magnitudes with localized, persistent error bands that propagate in time, suggesting that pointwise (strong-form) forcing corrections do not sufficiently constrain the evolving solution under repeated rollout. The weak-form correction yields a more uniformly bounded error field, with reduced peak values and fewer localized artifacts, consistent with improved numerical stability of the learned correction. Overall, the temporal error curves (Figure~\ref{Fig_relativeError_1Dconv}) and space-time diagnostics (Figure~\ref{Fig_Error_field_1Dconv}) indicate that weak-form correction is better aligned with the discretized convection-diffusion dynamics and delivers more robust improvements in multistep forecasting.\\
\indent Table~\ref{tab:timeavg_relL2_1DConvDiff} reports the time-averaged relative $L^2$ rollout error, $\overline{\varepsilon}_{L^2}$, along with the computational speedup relative to the high-resolution solver. The metric is defined as $\overline{\varepsilon}_{L^2}=\frac{1}{T}\int_{0}^{T}e(t)dt$. The low-resolution baseline ($u_L$) is faster than the high-resolution solver ($18.9\times$) but exhibits substantial accumulated error. Both correction strategies substantially improve accuracy while also maintaining similar speedups. The strong-form correction reduces the rollout error to $4.60\times 10^{-1}$, corresponding to a $3.26\times$ reduction in error relative to $u_L$, and achieves a $17.6\times$ speedup. The weak-form correction further lowers the error to $7.14\times10^{-2}$ (an additional $6.4\times$ reduction in error relative to the strong-form correction and $21 \times$ relative to $u_L$, while maintaining a similar speedup $17.5 \times$). Overall, the weak-form correction provides the best accuracy, delivering the lowest time-averaged rollout error at essentially the same runtime as the strong-form approach.
\begin{table}[t]
\centering
\caption[Time-averaged relative $L^2$ error and speedup]{%
TIME-AVERAGED RELATIVE $L^2$ ERROR OVER THE ROLLOUT INTERVAL $t\in[0,5000\Delta t]$ AND COMPUTATIONAL COST RELATIVE TO THE HIGH-RESOLUTION SOLVER.}
\label{tab:timeavg_relL2_1DConvDiff}
\footnotesize
\setlength{\tabcolsep}{4pt}
\renewcommand{\arraystretch}{1.15}

\begin{tabularx}{\columnwidth}{@{}>{\raggedright\arraybackslash}Xcc@{}}
\toprule
Method & $\overline{\varepsilon}_{L^2}$ & Speedup ($\times$) \\
\midrule
High-resolution reference ($u_H$) & 0.0 & 1.0 \\
Low-resolution baseline ($u_L$)        & $1.5$ & $ \text{18.9} $ \\
Strong-form correction                  & $4.60\times 10^{-1}$ & $ \text{17.6} $ \\
Weak-form correction                    & $7.14\times 10^{-2}$ & $ \text{17.5} $ \\
\bottomrule
\end{tabularx}
\end{table}
\subsection{Two-dimensional flow past a cylinder}
\begin{figure}[t]
  \centering
   \includegraphics[scale=0.9,width=0.9\textwidth]{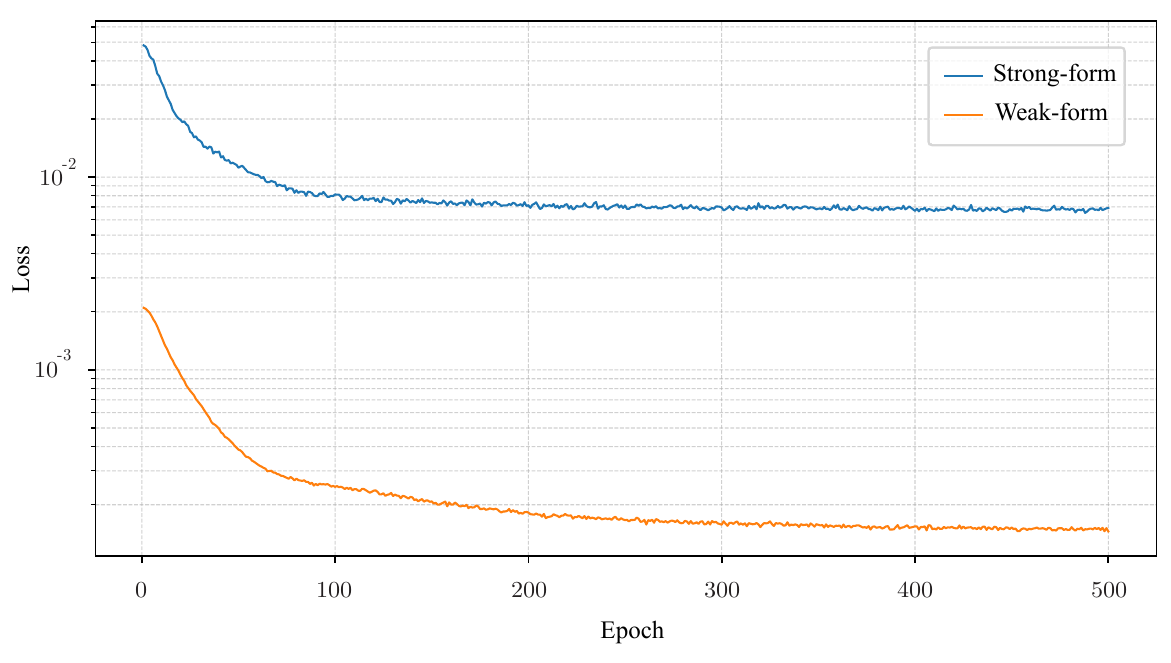}
  \caption{TRAINING LOSS VERSUS EPOCH FOR THE STRONG-FORM AND WEAK-FORM CORRECTIONS APPLIED TO 2D FLOW PAST A CYLINDER. }\label{Fig_loss_field_2Dcylinder}
\end{figure}
\begin{figure}[t]
  \centering
   \includegraphics[scale=0.9,width=0.9\textwidth]{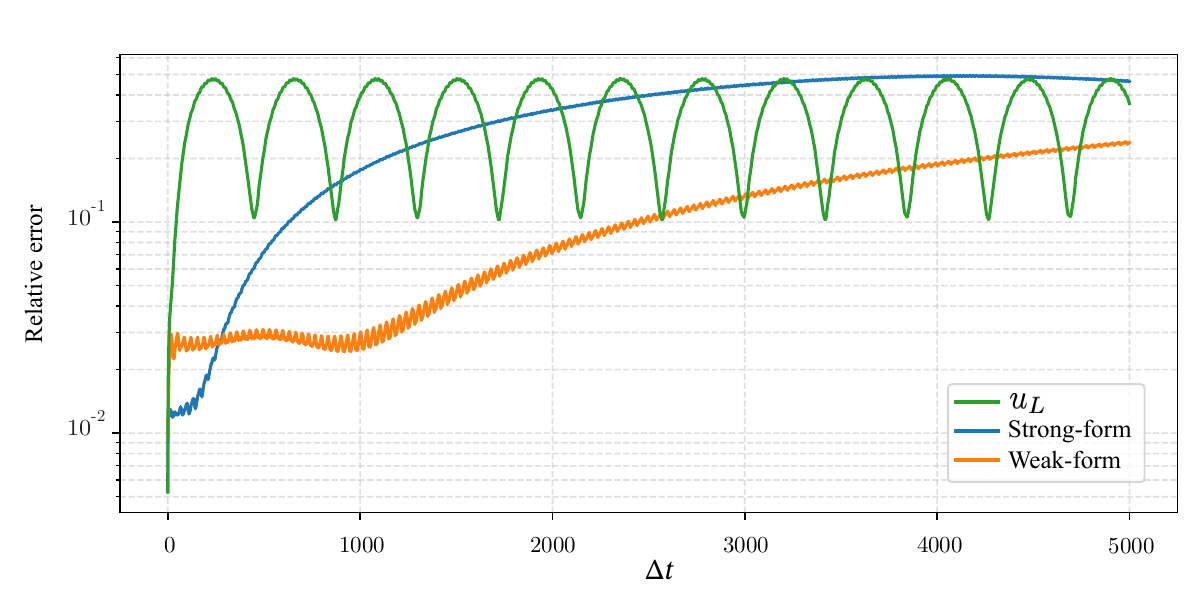}
  \caption{TIME EVOLUTION OF THE RELATIVE ROLLOUT ERROR FOR THE 2D FLOW PAST A CYLINDER USING STRONG-FORM AND WEAK-FORM NEURAL CORRECTIVE OPERATORS.}\label{Fig_rel_error_2Dcylinder}
\end{figure}
Following Section~\ref{subsec:methodology_weak-form_correction}, we train the learned
corrective operators using a CN Oseen-type weak form.
Let $\delta_t \mathbf{u}_h^{n+1}:=(\mathbf{u}_h^{n+1}-\mathbf{u}_h^{n})/\Delta t$ with
$\Delta t=10^{-2}$ and $\theta_{CN}=0.5$. We seek
$(\mathbf{u}_h^{n+1},p_h^{n+1})\in V_h\times Q_h$ such that, for all
$(\mathbf{v},q)\in V_h\times Q_h$,
\begin{equation}\label{eq:cn-weakcorr}
\begin{aligned}
(\delta_t\mathbf{u}_h^{n+1},\mathbf{v})_\Omega
&+\theta_{CN}\,\mathcal{A}(\mathbf{u}_h^{n+1},p_h^{n+1};\mathbf{v},q) \\
&+(1-\theta_{CN})\,\mathcal{A}(\mathbf{u}_h^{n},p_h^{n};\mathbf{v},q)=0 .
\end{aligned}
\end{equation}
The Oseen operator is
\begin{equation}\label{eq:operator-A}
\begin{aligned}
\mathcal{A}(\mathbf{u},p;\mathbf{v},q)={}&
\big((\nu+\nu_t)\nabla\mathbf{u},\nabla\mathbf{v}\big)_\Omega \\
&+\big(1+c_{\mathrm{adv}}\big)\,
c_{\mathrm{skew}}\!\left(\mathbf{u}_{\mathrm{adv}}^{n};\mathbf{u},\mathbf{v}\right) \\
&-(p,\nabla\!\cdot\!\mathbf{v})_\Omega
+(q,\nabla\!\cdot\!\mathbf{u})_\Omega \\
&+\big(\gamma\,\nabla\!\cdot\!\mathbf{u},\nabla\!\cdot\!\mathbf{v}\big)_\Omega
+\varepsilon_p (p,q)_\Omega .
\end{aligned}
\end{equation}
We use $\nu=10^{-3}$ and a small pressure regularization
$\varepsilon_p=10^{-10}$ for numerical stability.\\
\indent We learn three corrective coefficient fields ($c_{\mathrm{adv}}$, $\nu_t$, $\gamma$), introduced in Section~\ref{subsec:methodology_weak-form_correction}. At each CN step, these fields are predicted by a neural network from the current velocity state, 
\begin{equation} \big(c_{\mathrm{adv}},\nu_t,\gamma\big) = \mathcal{N}_\theta(\mathbf{u}_h). 
\end{equation} 
The learned corrective fields are discretized in a lower-order finite element space with $N_{\mathrm{coef}}=4732$ DOFs for $P=2$ and two velocity components. The neural network takes as input the current velocity field represented in the same scalar space and outputs the three coefficient fields $(c_{\mathrm{adv}}, \nu_t, \gamma)$, giving an output dimension of $3N_{\mathrm{coef}}$. We train by minimizing a multistep rollout loss over $m=20$ time steps. The dataset spans the time interval $t \in [0,53]$, including 300$\Delta t$ for training and 5000$\Delta t$ for testing. Training is performed for 500 epochs with mini-batch size $B=10$. Optimization uses Adam with an initial learning rate $10^{-4}$ and an exponential schedule per epoch, with a minimum learning rate of $10^{-6}$. The corrective operator $\mathcal{N}_\theta$ is implemented as a fully connected multilayer perceptron with three hidden layers of widths $(5000,4000,3000)$.\\ 
\indent The training loss histories for the strong-form and weak-form corrections are compared in Figure~\ref{Fig_loss_field_2Dcylinder}. Both approaches display a typical two-stage optimization trend, with a sharp reduction over the first 100 epochs followed by a slower, asymptotic decay. Importantly, the weak-form correction maintains a substantially lower loss throughout training. It starts at $\mathcal{O}(10^{-3})$, drops rapidly during the early stage, and continues to decrease gradually to $\mathcal{O}(10^{-4})$ by 500 epochs. In contrast, the strong-form loss begins at $\mathcal{O}(10^{-2})$ and plateaus near $\sim 7\times 10^{-3}$ after the initial decay, showing only small fluctuations thereafter. Overall, Figure~\ref{Fig_loss_field_2Dcylinder} indicates that weak-form training reaches a markedly lower terminal value (roughly $40$--$50\times$ lower than the strong-form case). From a numerical perspective, this behavior is consistent with the weak-form formulation providing a better-conditioned enforcement of the governing constraints via integral residuals, which are less sensitive to localized high-frequency discrepancies and can yield smoother, more stable optimization dynamics.\\
\begin{figure}[!htp]
  \centering
   \includegraphics[scale=0.9,width=0.9\textwidth]{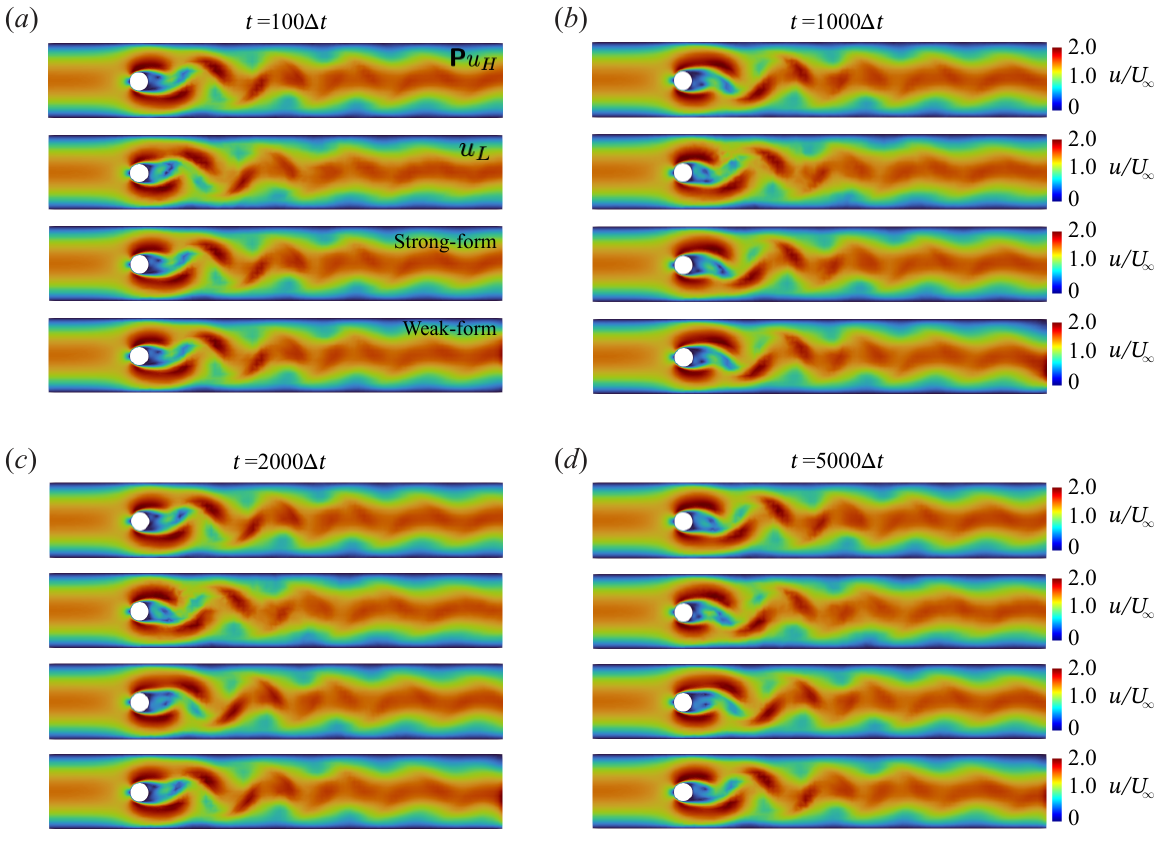}
  \caption{SNAPSHOTS OF ROLLOUT PREDICTIONS AT $t$ = [$100\Delta t$, $1000\Delta t$, $2000\Delta t$, $5000\Delta t$] USING STRONG-FORM AND WEAK-FORM NEURAL CORRECTIVE OPERATORS.}\label{Fig_Snapshots_rollout_predictions}
\end{figure}
\begin{figure}[!htp]
  \centering
   \includegraphics[scale=0.9,width=0.9\textwidth]{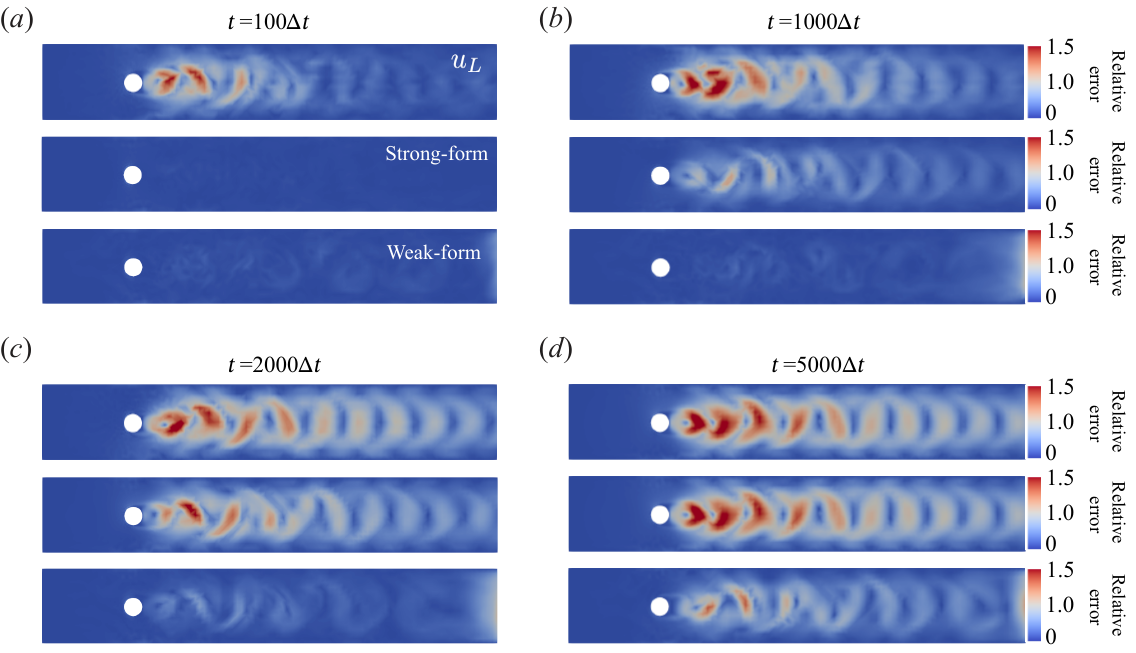}
  \caption{SNAPSHOTS OF ROLLOUT RELATIVE ERRORS AT $t = [100\Delta t,\ 1000\Delta t,\ 2000\Delta t,$ $5000\Delta t]$ USING STRONG-FORM AND WEAK-FORM NEURAL CORRECTIVE OPERATORS.}\label{Fig_Snapshots_rollout_relative_errors}
\end{figure}
\indent Figure~\ref{Fig_rel_error_2Dcylinder} shows the rollout relative error of Eq.~(\ref{eq:relativeerror}) as a function of time step ($\Delta t$), which we use to evaluate the trained model for corrected dynamical simulations. The baseline low-fidelity solution $u_{L}$ exhibits pronounced, periodic errors of $\mathcal{O}(10^{-1})$, indicating phase and amplitude mismatches and the accumulation of time-integration error in the vortex-shedding regime. This behavior is also visible in the second-row ($u_{L}$) of snapshots in Figures~\ref{Fig_Snapshots_rollout_predictions} and~\ref{Fig_Snapshots_rollout_relative_errors}. Both corrected models reduce the short-term error relative to $\tilde{u}_{L}$, as shown in the second-row (strong-form) and third-row (weak-form) results in Figure~\ref{Fig_Snapshots_rollout_relative_errors}. However, their long-horizon behavior differs substantially, as illustrated in Figure~\ref{Fig_Snapshots_rollout_relative_errors}(\textit{b})–(\textit{d}).\\
\indent Table~\ref{tab:timeavg_relL2_2Dcylinder} reports the time-averaged relative $L^{2}$ error for both short- ($t\in[0,1000\Delta t]$) and long-horizon ($t\in[0,5000\Delta t]$) rollouts, along with the associated computational cost. The low-resolution baseline $u_{L}$  exhibits a large mean error $\mathcal{O}(10^{-1})$ on both time intervals. Both correction strategies markedly reduce the short-horizon error on $t \in [0,1000\Delta t]$. The strong-form correction decreases $\overline{\varepsilon}_{L^{2}}$ from $3.07 \times 10^{-1}$ to $7.78 \times 10^{-2}$ (about $4 \times$ improvement) with essentially unchanged computational cost ($158 \times$ speedup). The weak-form correction yields the best short-term performance, achieving $2.73 \times 10^{-2}$ (about $11 \times$ improvement over $u_{L}$) while retaining a comparable speedup ($156\times$). Over the longer horizon $t \in [0,5000\Delta t]$, the two corrected models separate more clearly. The strong-form correction loses its advantage and degrades to $3.39 \times 10^{-1}$, slightly worse than the baseline, suggesting error accumulation and reduced long-term stability. In contrast, the weak-form correction maintains a substantially lower mean error $1.10 \times 10^{-1}$, demonstrating improved long-horizon robustness while preserving nearly the same acceleration as the low-resolution solver. For the vortex shedding frequency, the weak-form correction matches the high-resolution reference, with $St_{\text{weak}} \approx 0.195 \approx St_{u_H}$, whereas the low-resolution solver and strong-form correction underpredict it $\left(St_{u_L} \approx 0.188,\; St_{\text{strong}} \approx 0.186\right)$. In summary, the weak-form approach provides the best combination of accuracy and efficiency for corrected rollouts.

\begin{table}[t] 
\centering 
\caption{TIME-AVERAGED RELATIVE $L^2$ ERROR OVER THE ROLLOUT INTERVALS $t\in[0,1000\Delta t]$ AND $t\in[0,5000\Delta t]$ AND COMPUTATIONAL COST RELATIVE TO THE HIGH-RESOLUTION SOLVER.} \label{tab:timeavg_relL2_2Dcylinder} 
\footnotesize 
\setlength{\tabcolsep}{4pt} 
\renewcommand{\arraystretch}{1.15} 
\begin{tabularx}{\columnwidth}{@{}>{\raggedright\arraybackslash}Xccc@{}} \toprule 
Method & $\overline{\varepsilon}_{L^2}\,[0,1000\Delta t]$ & $\overline{\varepsilon}_{L^2}\,[0,5000\Delta t]$ & Speedup ($\times$) \\ \midrule 
High-resolution reference ($u_H$) & 0 & 0 & 1 \\ 
Low-resolution baseline ($u_L$) & $3.07\times 10^{-1}$ & $3.27\times 10^{-1}$ & 169 \\ 
Strong-form correction & $7.78\times 10^{-2}$ & $3.39\times 10^{-1}$ & 158 \\
Weak-form correction & $2.73\times 10^{-2}$ & $1.10\times 10^{-1}$ & 156 \\ \bottomrule 
\end{tabularx}
\end{table}


\section{Conclusion} 
We introduced a differentiable weak-form learning approach to accelerate incompressible finite element simulations by augmenting the momentum equation with learned bilinear operators assembled directly in the variational form. Implemented in Firedrake and trained end-to-end by using discrete adjoints coupled to PyTorch, the approach preserves key numerical structure of the underlying Galerkin discretization. Across the 1D convection--diffusion problem and 2D incompressible Navier--Stokes equations for flow past a cylinder at $Re=100$, weak-form correction consistently improved rollout accuracy and long-horizon stability relative to both the uncorrected low-resolution solver and an analogous strong-form correction, while maintaining comparable speedups of $\mathcal{O}(10^{1}-10^{2})$. In both cases, weak-form learning was more robust to error accumulation in long-horizon rollouts, yielding lower time-averaged relative $L^{2}$ error at essentially the same cost as the low-resolution simulation.\\ 
\indent Future work will evaluate the proposed approach on additional engineering metrics for the cylinder benchmark, including drag and lift coefficients. We will also explore extending the weak-form correction framework to other applications, assessing generalization across Reynolds numbers and geometries, and scaling training and inference to three-dimensional turbulent flows on unstructured meshes. Finally, we will explore alternative neural architectures for parameterizing the corrective operators, including graph neural networks. An executable reproducibility package (implementation, configurations, and evaluation scripts) is available for verification upon reasonable request.


\section*{Acknowledgments}
The work was supported by the Argonne Leadership Computing Facility (ALCF) Postdoctoral Fellowship and by the U.S. Department of Energy, Office of Science, Office of Advanced Scientific Computing Research (ASCR) and the Scientific Discovery through Advanced Computing (SciDAC) FASTMath Institute program and Competitive Portfolios Project on Energy Efficient Computing: A Holistic Methodology under Contract No. DE-AC02-06CH11357. JJ thanks Dr. Bethany Lusch for valuable discussions on differentiable training and the implementation of PyTorch.

\section*{Author ORCIDs}
\orcidlink{https://orcid.org} Junoh Jung \href{https://orcid.org/0000-0003-0962-3127}{https://orcid.org/0000-0003-0962-3127};\\
\orcidlink{https://orcid.org} Emil Constantinescu \href{https://orcid.org/xxx}{https://orcid.org/0000-0002-7003-6899}.

\begin{appendices}

\end{appendices}

\bibliographystyle{plainnat}
\bibliography{arXiv_template}

 \begin{center}
	\scriptsize \framebox{\parbox{5in}{Government License (will be removed at publication):
			The submitted manuscript has been created by UChicago Argonne, LLC,
			Operator of Argonne National Laboratory (``Argonne").  Argonne, a
			U.S. Department of Energy Office of Science laboratory, is operated
			under Contract No. DE-AC02-06CH11357.  The U.S. Government retains for
			itself, and others acting on its behalf, a paid-up nonexclusive,
			irrevocable worldwide license in said article to reproduce, prepare
			derivative works, distribute copies to the public, and perform
			publicly and display publicly, by or on behalf of the Government. The Department of Energy will provide public access to these results of federally sponsored research in accordance with the DOE Public Access Plan. http://energy.gov/downloads/doe-public-access-plan.
}}
	\normalsize
\end{center}

\end{document}